\documentclass[oribibl]{llncs}

\usepackage{url}
\usepackage{amsmath}
\usepackage{balance}
\usepackage{natbib}
\usepackage{booktabs}
\usepackage{tabularx}
\usepackage{graphicx}
\usepackage{threeparttable}
\usepackage{natbib}
\bibpunct[, ]{(}{)}{;}{a}{}{,}

\usepackage[colorlinks=true,linkcolor=cyan,
  citecolor=cyan,urlcolor=cyan]{hyperref}

\usepackage{fancyhdr} % for arxiv
\fancypagestyle{firststyle}
{
   \fancyhf{}
   \fancyfoot[C]{\scriptsize{SN Social Sciences, 2024, vol.~4, article no.~232, pp.~1--18. This version is the author's copy. The publisher's definite version is available via \url{https://doi.org/10.1007/s43545-024-01029-x}.}}
}

\begin{document}

\title{A Comparative Study of Online \\ Disinformation and Offline Protests}

\author{Jukka Ruohonen\orcidID{0000-0001-5147-3084} \email{juk@mmmi.sdu.dk}}
\institute{University of Southern Denmark, S\o{}nderborg, Denmark}

\maketitle
\thispagestyle{plain}

\begin{abstract}
  This paper evaluates the effect of online disinformation upon offline
  political protests with a time series cross-sectional sample of $125$
  countries in a period between 2000 and 2019. The results are mixed. Based on
  Bayesian multi-level regression modeling, (i)~there indeed is an effect
  between online disinformation and offline protests, but the effect is
  partially meditated by political polarization. The results are clearer in a
  sample of countries belonging to the European Economic Area. With this sample,
  (ii)~offline protest counts increase from online disinformation disseminated
  by domestic governments, political parties, and politicians as well as by
  foreign governments. Furthermore, (iii)~Internet shutdowns tend to decrease
  the counts, although, paradoxically, the absence of governmental online
  monitoring of social media tends to also decrease these. With these results,
  the paper contributes to the blossoming disinformation research by modeling
  the impact of disinformation upon offline phenomenon. The contribution is
  important due to the various policy measures planned or already enacted.
\vspace{5pt}
\end{abstract}

\section{Introduction}

\thispagestyle{firststyle} % for arxiv

In early 2021 the United States Capitol in Washington was stormed during a riot
and violent attack. A similar storming occurred in Brazil in 2023. Although both
attacks were instances in longer sequences of events, these have provided a
testimony for many observers who had claimed that online actions, including the
propagation of disinformation, have offline consequences. Soon after, a number
of papers have been published about the relation between online disinformation
and offline violence, among other related relations.

Recently, an intriguing study was published on the relationship between online
disinformation and domestic terrorism. According to the paper's result,
disinformation spread by governmental actors and political parties contributes
to domestic terrorism but the effect is mediated by political polarization and
tribalism within countries~\citep{Piazza21}. According to rather similar results
in a different paper, discussions on social media between opposing groups are
associated with offline physical violence~\citep{Gallacher21}. These results
further align with those showing that disinformation spread by foreign actors is
associated with physical violence \citep{Arayankalam21}. A relation has been
also found between online hate speech and offline hate
crimes~\citep{Williams20}. To this end, it has also been observed that white
supremacist online propaganda tends to later appear offline in the form of
fliers, banners, and graffiti~\citep{Diab23}. Closer to the present study, it
has been observed that exposure to online news is positively associated with
participation in political demonstrations, including particularly in
authoritarian countries \citep{Kirkizh21}. While there is thus a solid
foundation for the present work, thus far, the relation between online
disinformation and offline political protests has not been explored. This gap in
existing research is surprising already due to the studies exploring the
relation of online phenomena to violence, which is a much stronger presumption
than a mere political protest.

It has been widely observed that the Internet and social media increase
political activity due to the ease of exchanging information and recruiting new
members for a common cause, emotional and motivational appeals, strengthening of
group identities, agenda-setting and empowerment, and related
factors~(\citealt{Brennan18}; \citealt{Brunsting02}; \citealt{Jost18};
\citealt[pp.~22--36]{Schumann15}; \citealt{Willnat13}). The perhaps most often
cited example is the Arab Spring in the 2010s during which social media was
successfully exploited for uprisings and other political purposes. According to
empirical results, indeed, coordinated posts in social media increased the
volume of protests the next day~\citep{SteinertThrelkeld15}. Though,
paradoxically, despite all the bells and whistles, there are also studies
indicating that maintaining social media presence may decrease political
participation~\citep{Theocharis16}. In terms of social movements, keeping an
online movement alive for offline action is always a
challenge~\citep{Bu17}. According to skeptical viewpoints, the weak ties of
online engagement lack power and seldom lead to new political
possibilities~\citep{Unver17}. If nothing else, these critical results and
skeptical viewpoints serve to underline that much still remains unknown about
online politics and their relation to offline activities.

The same applies to disinformation research~\citep{Guess20}. Here, too, again,
the Arab Spring serves as an important but sombre historical milestone. The
early enthusiasm about the Internet's democratizing power soon after changed
into a disillusionment. First came ISIS, then came Brexit, and later came the
2016 and 2020 presidential elections in the United States.\footnote{~Islamic
  State of Iraq and the Levant (ISIS).} All four examples are also important
landmarks of disinformation and its research. As is soon further discussed in
Section~\ref{sec: framing}, the examples also underline a state-centric
viewpoint to disinformation; ISIS was skillful with propaganda, as was Russia
with its alleged foreign election interference in the United States and
elsewhere. Although it has been questioned how successful Russia was in
  its past election interference endeavors~\citep{Eady23}, it can be still
  acknowledged that at least they were skillful. Also other actors---from
political parties to charlatans---quickly learned how the online disinformation
game is played in the current algorithmically curated and platform-dominated
information ecosystems~\citep{Bradshaw20}. Given this brief motivating
introduction, the paper examines the following hypothesis:

\begin{quote}
\textit{A hypothesis: online disinformation disseminated by governments,
  political parties, and politicians, whether to domestic or foreign audiences,
  is associated with offline political protests, either decreasing or increasing
  the frequency of these, possibly via mediation with other political phenomena
  such as political polarization.}
\end{quote}

For examining the hypothesis, the paper proceeds in a straightforward manner. To
further sharpen the hypothesis, the paper's framing is first elaborated in more
detail in Section~\ref{sec: framing}. Thereafter, the research design is
discussed in Section~\ref{sec: research design}, the empirical results are
presented in Section~\ref{sec: results}, and the conclusion follow in the final
Section~\ref{sec: conclusion}.

\section{Framing}\label{sec: framing}

The online and offline realms increasingly intervene, and this poses major
challenges to democracy and civil rights~\citep{Ruohonen21MIND,
  Unver17}. Disinformation is among these challenges. In the past---not so long
ago, a propagandist eager to spread disinformation had to have a printing press
to publish books containing false information, an airplane from which to drop
leaflets, or a television channel via which disinformation could be broadcast to
masses. But those days are long gone. The Internet and its algorithms make
propagation of disinformation easy.

A few words must be written about terminology, which is still debated in
academia and elsewhere \citep{Altay23, Choras21, Guess20, Horne21}. There is a
good reason for these debates; it is difficult to make demarcations between
disinformation, misinformation, propaganda, fake news, hoaxes, conspiracy
theories, satire, memes, deep fakes, and whatever else the Internet and its
subcultures continuously invent---together with intelligence agencies and other
governmental bodies. Nevertheless, to make at least some sense out of the
terminological confusion, misinformation can be defined as a unintentional
adaption or dissemination of false or inaccurate information; conspiracy
theories and related phenomena belong to this category. Disinformation, in turn,
can be seen to involve intentional propagation of information that is known to
be false or misleading in order to reach specific goals. Here, intention is the
keyword; disinformation is a weapon particularly in the hands of governmental
bodies and political actors. But, then, so is propaganda, which, however, can be
conceptually separated from disinformation because propaganda may also deliver
information that is true. While keeping these points in mind, the dataset used
defines disinformation simply as ``misleading viewpoints or false information.''
Given that this simple characterization may also refer to misinformation,
propaganda, or something else, the choice to prefer the term disinformation is
somewhat misleading but can be justified on the grounds that the theoretical
hypothesis presumes an intention to provoke offline political protests. For any
empirical purposes, the choice of a term does not matter.

Furthermore, the term disinformation should be, in the present context, prefixed
with the term state-centric. There are three reasons for this
additional qualifying term, which generally refers to actions and policies of
states in a contrast to actions and policies of other actors, including private
sector actors, international but not multilateral organizations,
non-governmental organizations, and related actors operating in the online
environment. The first reason is practical; the dataset used specifically
includes only variables on disinformation that is disseminated by actors
explicitly tied to countries' political systems. This framing aligns
with other recent studies that have examined counter-propaganda efforts by
state-affiliated organizations~\citep{Ruohonen21GOODIT}. The second reason
originates from the comparative research approach pursued; here, the term
state-centric is often used to distinguish research operating with inter-state
relations in a contrast to multi-level or global
perspectives~\citep{Ebbinghaus98}. The third and final reason relates to the
dependent variable. As is soon clarified in Section~\ref{sec: dependent
  variable}, the disinformation variables are regressed against a variable that
measures protests against a state or its policies. Together, these
terminological clarifications also sharpen the underlying hypothesis.

The disinformation and propaganda disseminated by ISIS and Russia have had one
thing common; the primary (but not the only) target has been Western
democracies, and the intention has been to deepen the existing societal
cleavages and sabotage the internal cohesion within Western countries by
exploiting weaknesses in the current engagement economy and conducting violent
offline attacks, whether via terrorism or assassinations~\citep{Dundon22,
  Hamilton19, Kalniete21, Rosenblatt19, Sultan19, Watling24,
  ZhangLukito21}. Hate is a powerful method, whether in the hands of Muslim
minorities or right-wing extremists.

To reach the goals, which, as said, are a distinct trait of disinformation, time
and sufficient resources are needed. In a sense, one still metaphorically needs
the printing press, the airplane, or the television channel; to exploit societal
weaknesses, one needs to be well-educated. The same applies when the goal of the
exploitation is to provoke political protests in foreign countries. Education,
staffing, and resources are thus another factor separating state-centric
disinformation from other false information propagated by laypeople. Though: if
the online and offline realms intervene, so do the old logic and the new logic
of media; many national disinformation outlets, some of which are linked to
foreign disinformation organizations, rely on national media for their criticism
and distortion of narratives~\citep{Toivanen21}. For building such
disinformation outlets, nevertheless, time is needed even when the result would
be amateurish due to lack of education.

Time is important also methodologically; many of the studies exploring the
online-offline nexus have relied on cross-sectional data or survey snapshots,
although a longitudinal focus too is necessary. Finally, time, resources, and
skill are needed also by political parties and politicians who rely on false
information in order to reach political goals via the emotionally driven
information ecosystems~\citep{Ruohonen20MISDOOM}. To conclude: the framing
described aligns with the longitudinal cross-country analysis pursued. The
design for the analysis is thus subsequently elaborated.

\section{Research Design}\label{sec: research design}

\subsection{Data}

Three datasets are used for the empirical analysis. The first is the mass
protest dataset assembled by \citet{Massmob16a}. It provides the dependent
variable for the analysis. The primary independent variables for disinformation
are based on the celebrated \text{V-Dem} dataset~\citep{VDem21a}. Also the
secondary independent variables come from the same dataset. Finally, two control
variables were retrieved from \citeauthor{WorldBank21b}'s
\citeyearpar{WorldBank21b,WorldBank21d} online data portal. The datasets were
merged by including only those countries that were present in all datasets with
no missing values. Although merging reduced the number of countries---from $162$
in the protest dataset to $n = 125$ in the sample used, the absence of missing
values outweigh the reduction and the use of interpolation techniques. It is
also worth noting that, for a reason or another, the protest dataset misses the
United States, which is a clear limitation in the present disinformation
context. Given that there are $t = 20$ years for each country, covering a period
from 2000 to 2019, the sample size is still more than sufficient for statistical
analysis.

\subsection{Dependent Variable}\label{sec: dependent variable}

\begin{figure}[th!b]
\centering
\includegraphics[width=\linewidth, height=5cm]{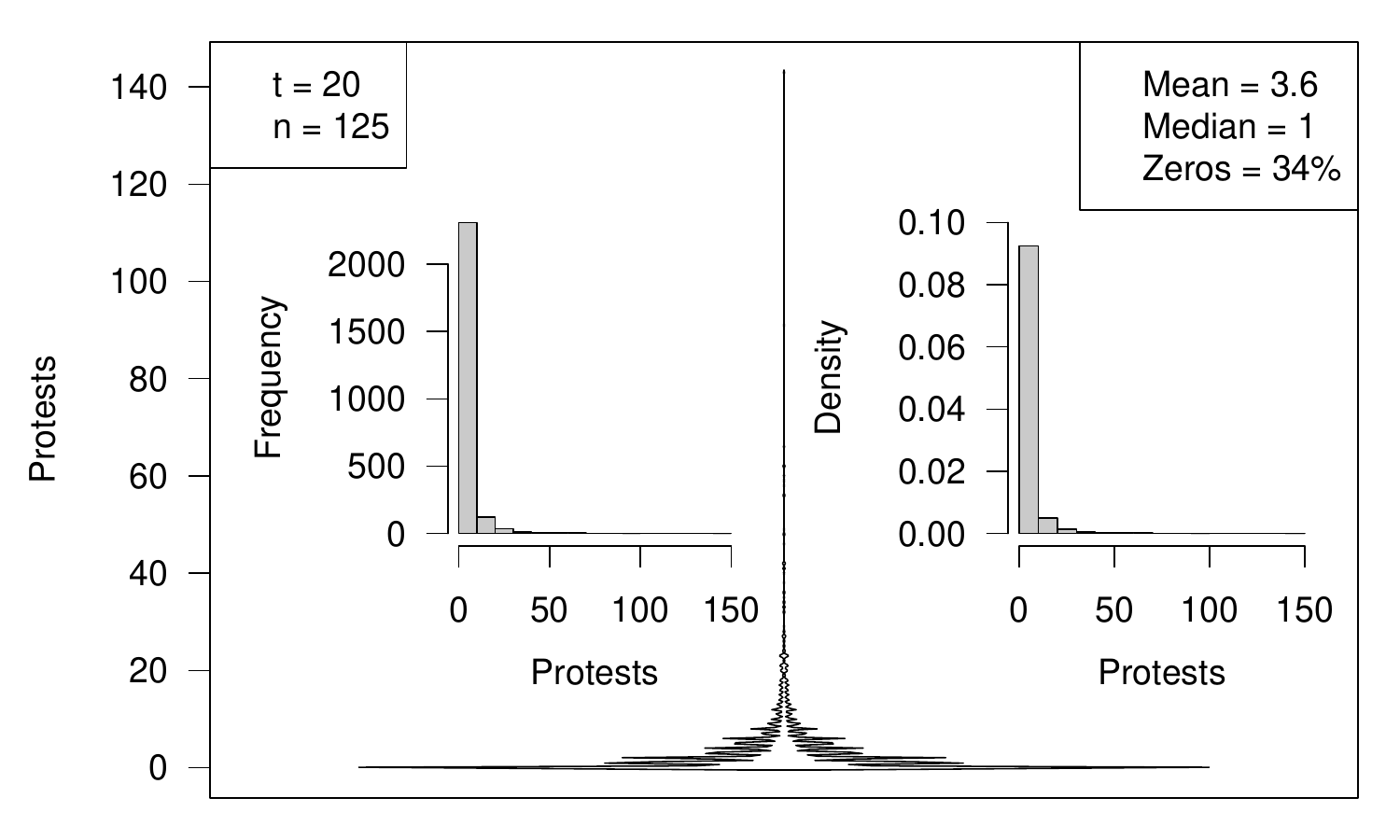}
\caption{The Dependent Variable: Protest Counts}
\label{fig: protests}
\end{figure}

The protest variable measures a gathering of at least fifty people who oppose a
state or its policy in a given country. As clarified by \citet{Massmob16b}, the
variable excludes (a)~protests opposing a foreign state or a group of states as
well as (b)~demonstrations with non-state targets, including socioeconomic,
religious, or other groups protesting rival groups---even when these have
required intervention from police forces. Moreover, (c) industrial action only
counts as a protest if people protest outdoors for a state-level labor policy
instead of protesting against a company or an employer association. Finally,
(d)~armed resistance and rebel groups are excluded. All in all, the dependent
variable thus measures conventional political protests against a state or the
policies it has enacted or is about to enact. As for operationalization, the
individual protests reported in the dataset were aggregated into annual
counts. As can be seen from the outer beanplot in Fig.~\ref{fig: protests}, and
the inner histogram and density plot, the result is a typical count data
variable; there are many country-year pairs with no protests but the
distribution has a long right tail.

\subsection{Independent Variables}\label{sec: independent variables}

The primary independent variables are obviously related to disinformation. There
are five variables in the \text{V-Dem} dataset for disinformation.\footnote{~The
  names of the original variables are: \texttt{v2smgovdom}, \texttt{v2smgovab},
  \texttt{v2smpardom}, \texttt{v2smparab}, and \texttt{v2smfordom}.} All of
these have been used in recent research~\citep{Arayankalam21, Piazza21}, and all
of these are used also in the present work.

The first two measure whether and how often a government, its agencies, and
agents working on its behalf disseminate disinformation on social media for
domestic and foreign audiences. These two variables on governmental
disinformation dissemination are ideal for examining the state-centric
viewpoint. Although neither variable considers the context and content of the
disinformation spread, the distinction between domestic and foreign audiences
aligns with war propaganda, which is often exported to foreign consumption and
designed for domestic audiences in order to counter war weariness and related
concerns~\citep{Ruohonen21GOODIT}. Certainly, disinformation is propagated
online also by numerous other non-governmental entities; fortunately, therefore,
\text{V-Dem} provides also analogous two variables regarding the frequency of
disinformation distributed to domestic and foreign audiences by major political
parties and political candidates. Again, the distinction between domestic and
foreign audiences is important because political parties in one country may
disseminate false information to another country~\citep{Ruohonen20MISDOOM}. The
Internet makes such dissemination easy. The fifth variable is about imported
disinformation; whether a foreign government and its lackeys have targeted a
given country with their exported disinformation. Needless to say, the alleged
election interference by Russia and other countries provides a solid rationale
also for this variable.

Six independent variables are included for capturing the policy-side. A few
examples are in order to contextualize this side. Although heavily
criticized~\citep{Schultz20} and now deprecated, the Network Enforcement Act
(NetzDG) in Germany is a good example of legislative attempts to filter online
hate speech. Likewise, in 2018, a comparable law was passed in France; it allows
public authorities to remove fake information or even block sites that publish
such content~\citep{Nagasako20}. In addition to these national laws, the
European Union has been actively enacting different legislative measures and
informal codes of conduct, including for disinformation and hate
speech~\citep{Buiten20, Pataki21}. On the side of even more drastic measures,
particularly authoritarian countries in conflict zones have increasingly
resorted to Internet shutdowns; among the memorable historical examples was the
2011 shutdown of the Internet in Egypt in the face of the Arab Spring
uprisings~\citep{Arnaudo13}. Thus, there is a well-founded rationale to consider
also the policy-side variables.

The six variables are: (1)~the capacity of a government to filter online
content; (2) governmental filtering of online content in practice; (3) the
capacity of a government to shutdown national parts of the Internet; (4)
governmental alternatives to global social media platforms; (5) governmental
monitoring of social media; and (6) regulation of online
content~\citep[pp.~317--320]{VDem21b}. All variables are standardized akin to
$z$-values; a value zero approximates the mean of all country-year pairs in the
\text{V-Dem} dataset. For the capacity-related variables (1) and (3),
  higher values indicate a higher capacity of the measures taken. For the
  variables (2) and (5), higher values indicate that a government neither
  censors online political information nor monitors social media for political
  content. For the variable (4), higher values indicate that people seldom use
  state-controlled social media platforms. Finally, for the variable (6), higher
  values indicate that a government can only remove content based on
  well-defined legal criterion.

To align the present work with \citeauthor{Piazza21}'s \citeyearpar{Piazza21}
paper, an independent variable is included for political polarization. Unlike
measures and conceptualizations based on political parties and national party
systems~\citep{Lauka18, Sartori76}, the political polarization variable in the
\text{V-Dem} dataset is a composite variable measuring ``the extent to which
political differences affect social relationships beyond political
discussions,'' such that in polarized societies ``supporters of opposing
political camps are reluctant to engage in friendly interactions, for example,
in family functions, civic associations, their free time activities and
workplaces'' \citep[p.~224]{VDem21b}.\footnote{~The names of the independent
  variables in the \text{V-Dem} dataset are: \texttt{v2smgovfilcap},
  \texttt{v2smgovfilprc}, \texttt{v2smgovshutcap}, \texttt{v2smgovsmalt},
  \texttt{v2smgovsmmon}, \texttt{v2smregcon}, and \texttt{v2cacamps}, in the
  order of listing.}  Although the United States is not present in the sample
used, the country is currently the best known example of such society-wide
political polarization.

\subsection{Control variables}

Four control variables are used: a change in a country's governing regime type,
the presence and consumption of national online media, a country's gross
domestic product (GDP) in current prices~\citep{WorldBank21b}, and the total
population in a country~\citep{WorldBank21d}. The last two control variables
were used also by \citet{Piazza21}.

The first control variable is a re-coded dummy variable indicating whether a
country's past regime changed for any reason in a given year, whether due to a
loss in a war, a \textit{coup~d'\'etat}, an assassination, a popular uprising, a
civil war, a democratization process, a foreign interference, or something
analogous~\citep[p.~134]{VDem21b}. If the current regime is still in place, the
dummy variable takes a value zero. Although the dummy variable is included only
as a control variable, such that no theoretical interpretation is provided for
its potential statistical effect, there is still a rationale for its inclusion;
it is highly probable that a regime change leads to protests and increased
political activity both online and offline. The final control variable is the
existence of national online media and its consumption by domestic
audiences.\footnote{~The two control variables are named \texttt{v2smonex} and
  \texttt{v2regendtype} in the \text{V-Dem} dataset, respectively.}

\subsection{Methods}

The dependent variable represents the count of protests in a given country in a
given year. Given that no protests occurred in about one-third of the
country-year pairs, a conventional zero-inflated Poisson model is used for
statistical estimation of the protest counts. In general, the model is a standard Poisson regression that additionally takes into account the overdispersion caused by the zeros.

Unlike \citet{Arayankalam21}, who only consider cross-sectional estimates in a
single year, the estimation is carried in a time series cross-sectional (TSCS)
context, which is a natural choice due to the underlying panel data. The TSCS
context offers some clear benefits. Among these is the sample size; in total, $n
\times t = 20 \times 125 = 2500$ observations are present. Another benefit is
the possibility to evaluate and control the longitudinal dynamics together with
the cross-sectional dimension. There are also specific statistical estimators
that take the longitudinal dimensions into account~\citep{Arellano91}. Instead
of considering such dynamic estimators, however, the model specifications are
kept as simple as possible due to the large number of variables; the estimation
is carried out with so-called random effects models in which the cross-sectional
and annual effects are modeled with varying intercepts. In other words, there
are intercepts for both countries and years, and these are modeled as random
variables. This simple specification is well-known and
well-documented~\citep{Wooldridge10}. As the interest is only to control and not
interpret the annual and cross-sectional effects, the random effects model is
preferable---insofar as the effects are statistically
relevant~\citep{Antonakis19}. As for computation, Bayesian estimation is used
with the \texttt{brms} package~\citep{brms}, which provides a user friendly R
interface for the Stan machinery.

Three models are estimated. The first includes the five governmental
disinformation variables; the second the two party disinformation variables; and
the third all of the variables. The four control variables are included in all
models. Following the arguments that comparative research should focus on
different regimes, alliances, and unions of countries~\citep{Ebbinghaus98}, the
models are further recomputed with a subset of $26$ countries belonging to the
European Economic Area (EEC). This additional computation provides a good
re-check of the results because the EEC region is well-defined instead of being
a pseudo-random sample of countries in the world. Finally, interactions between
the variables are briefly considered, as in previous research~\citep{Piazza21,
  Arayankalam21}, but these are described later alongside the analysis.

\section{Results}\label{sec: results}

The multi-level regression results are summarized Tables~\ref{tab: reg
  results}~and~\ref{tab: reg results eec}. These can be disseminated with three
points. Before proceeding, it is worth remarking that the standard deviations
are fairly large for both the annual and the cross-sectional random effects. The
model specification thus seems adequate in this regard. The cross-sectional
variances are also larger than the annual variances. The overdispersion parameters too are clearly non-zero.

\begin{table*}[p!]
\centering
\caption{Regression Results: Full Sample (20 years, 125 countries)}
\label{tab: reg results}
\begin{tabular}{lrrrcrrrcrrr}
\toprule
& \multicolumn{3}{c}{Model 1.}
&& \multicolumn{3}{c}{Model 2.}
&& \multicolumn{3}{c}{Model 3.} \\
\cmidrule{2-12}
&& \multicolumn{2}{c}{95\% CI}
&&& \multicolumn{2}{c}{95\% CI}
&&& \multicolumn{2}{c}{95\% CI} \\
\cmidrule{3-4} \cmidrule{7-8} \cmidrule{11-12}
& Coef. & lower & upper
&& Coef. & lower & upper
&& Coef. & lower & upper \\
\hline
Intercept
& $-2.59$ & $-4.69$ & $-0.43$
&& $-2.69$ & $-4.83$ & $-0.53$
&& $-2.91$ & $-4.92$ & $-0.88$ \\
\cmidrule{1-1}
Gov.~disinformation: domestic
& $-0.17$ & $-0.26$ & $-0.09$
&&&&&& $-0.10$ & $-0.19$ & $0.00$ \\
Gov.~disinformation: abroad
& $-0.11$ & $-0.21$ & $-0.01$
&&&&&& $-0.10$ & $-0.21$ & $0.02$ \\
Foreign gov.~disinformation
& $-0.07$ & $-0.16$ & $0.02$
&&&&&& $-0.07$ & $-0.16$ & $0.02$ \\
\cmidrule{1-1}
Party disinformation: domestic
&&&&& $-0.13$ & $-0.21$ & $-0.05$
&& $0.02$ & $-0.07$ & $0.10$ \\
Party disinformation: abroad
&&&&& $-0.10$ & $-0.18$ & $0.00$
&& $0.06$ & $-0.05$ & $0.16$ \\
\cmidrule{1-1}
Gov.~filtering capacity
&&&&&&&&& $0.10$ & $0.00$ & $0.20$ \\
Gov.~filtering practice
&&&&&&&&& $0.05$ & $-0.05$ & $0.15$ \\
Gov.~shutdown capacity
&&&&&&&&& $-0.18$ & $-0.28$ & $-0.07$ \\
Gov.~social media alternatives
&&&&&&&&& $0.03$ & $-0.10$ & $0.16$ \\
Gov.~social media monitoring
&&&&&&&&& $-0.20$ & $-0.29$ & $-0.10$ \\
Regulation of online content
&&&&&&&&& $0.09$ & $-0.01$ & $0.19$ \\
Political polarization
&&&&&&&&& $0.16$ & $0.11$ & $0.21$ \\
\cmidrule{1-1}
National online media
& $0.01$ & $-0.05$ & $0.07$
&& $0.01$ & $-0.05$ & $0.07$
&& $-0.02$ & $-0.08$ & $0.05$ \\
Regime change
& $-0.03$ & $-0.13$ & $0.06$
&& $-0.01$ & $-0.11$ & $0.08$
&& $0.01$ & $-0.09$ & $0.11$ \\
$\ln(\textmd{population})$
& $0.35$ & $0.23$ & $0.48$
&& $0.35$ & $0.23$ & $0.47$
&& $0.36$ & $0.24$ & $0.48$ \\
$\ln(\textmd{GDP per capita})$
& $-0.25$ & $-0.33$ & $-0.16$
&& $-0.24$ & $-0.32$ & $-0.16$
&& $-0.23$ & $-0.31$ & $-0.14$ \\
\hline
Std.~dev.~years (intercepts)
& $0.29$ & $0.20$ & $0.42$
&&  $0.29$ & $0.20$ & $0.43$
&& $0.27$ & $0.18$ & $0.39$ \\
Std.~dev.~countries (intercepts)
& $1.07$ & $0.91$ & $1.24$
&& $1.00$ & $0.86$ & $1.16$
&& $0.99$ & $0.84$ & $1.17$ \\
\bottomrule
\end{tabular}
%\end{table*}
%
\vspace{10pt}
%
%\begin{table*}[th!b]
\centering
\caption{Regression Results: EEC Sample (20 years, 26 countries)}
\label{tab: reg results eec}
\begin{tabular}{lrrrcrrrcrrr}
\toprule
& \multicolumn{3}{c}{Model 1.}
&& \multicolumn{3}{c}{Model 2.}
&& \multicolumn{3}{c}{Model 3.} \\
\cmidrule{2-12}
&& \multicolumn{2}{c}{95\% CI}
&&& \multicolumn{2}{c}{95\% CI}
&&& \multicolumn{2}{c}{95\% CI} \\
\cmidrule{3-4} \cmidrule{7-8} \cmidrule{11-12}
& Coef. & lower & upper
&& Coef. & lower & upper
&& Coef. & lower & upper \\
\hline
Intercept
& $-4.87$ & $-14.05$ & $3.25$
&& $-4.33$ & $-12.22$ & $2.87$
&& $-2.55$ & $-14.12$ & $7.75$ \\
\cmidrule{1-1}
Gov.~disinformation: domestic
& $0.13$ & $-0.06$ & $0.32$
&&&&&& $0.47$ & $0.17$ & $0.76$ \\
Gov.~disinformation: abroad
& $-0.74$ & $-1.04$ & $-0.43$
&&&&&& $-0.61$ & $-0.99$ & $-0.21$ \\
Foreign gov.~disinformation
& $0.38$ & $0.18$ & $0.59$
&&&&&& $0.61$ & $0.34$ & $0.88$ \\
\cmidrule{1-1}
Party disinformation: domestic
&&&&& $0.07$ & $-0.11$ & $0.25$
&& $0.61$ & $0.34$ & $0.88$ \\
Party disinformation: abroad
&&&&& $-0.80$ & $-1.06$ & $-0.56$
&& $-0.62$ & $-0.96$ & $-0.28$ \\
\cmidrule{1-1}
Gov.~filtering capacity
&&&&&&&&& $0.12$ & $-0.34$ & $0.59$ \\
Gov.~filtering practice
&&&&&&&&& $-0.51$ & $-0.94$ & $-0.06$ \\
Gov.~shutdown capacity
&&&&&&&&& $-0.05$ & $-0.56$ & $0.48$ \\
Gov.~social media alternatives
&&&&&&&&&  $0.69$ & $0.19$ & $1.22$ \\
Gov.~social media monitoring
&&&&&&&&& $-0.38$ & $-0.65$ & $-0.10$ \\
Regulation of online content
&&&&&&&&& $-0.38$ & $-0.65$ & $-0.10$ \\
Political polarization
&&&&&&&&& $0.49$ & $0.28$ & $0.71$ \\
\cmidrule{1-1}
National online media
& $0.48$ & $0.30$ & $0.65$
&& $0.35$ & $0.17$ & $0.52$
&& $0.33$ & $0.12$ & $0.53$ \\
Regime change
& $-0.28$ & $-3.65$ & $3.10$
&& $-1.08$ & $-3.94$ & $1.79$
&& $-2.15$ & $-6.52$ & $2.15$ \\
$\ln(\textmd{population})$
& $0.47$ & $0.00$ & $1.00$
&& $0.59$ & $0.19$ & $1.05$
&& $0.45$ & $-0.16$ & $1.14$ \\
$\ln(\textmd{GDP per capita})$
& $-0.15$ & $-0.45$ & $0.15$
&& $-0.36$ & $-0.66$ & $-0.07$
&& $-0.28$ & $-0.60$ & $0.03$ \\
\hline
Std.~dev.~years (intercepts)
& $0.34$ & $0.23$ & $0.50$
&& $0.35$ & $0.23$ & $0.51$
&& $0.27$ & $0.18$ & $0.40$ \\
Std.~dev.~countries (intercepts)
& $1.63$ & $1.15$ & $2.28$
&& $0.35$ & $0.23$ & $0.51$
&& $2.07$ & $1.42$ & $3.06$ \\
\bottomrule
\end{tabular}
\end{table*}

\begin{figure*}[p!]
\centering
\includegraphics[width=\linewidth, height=18cm]{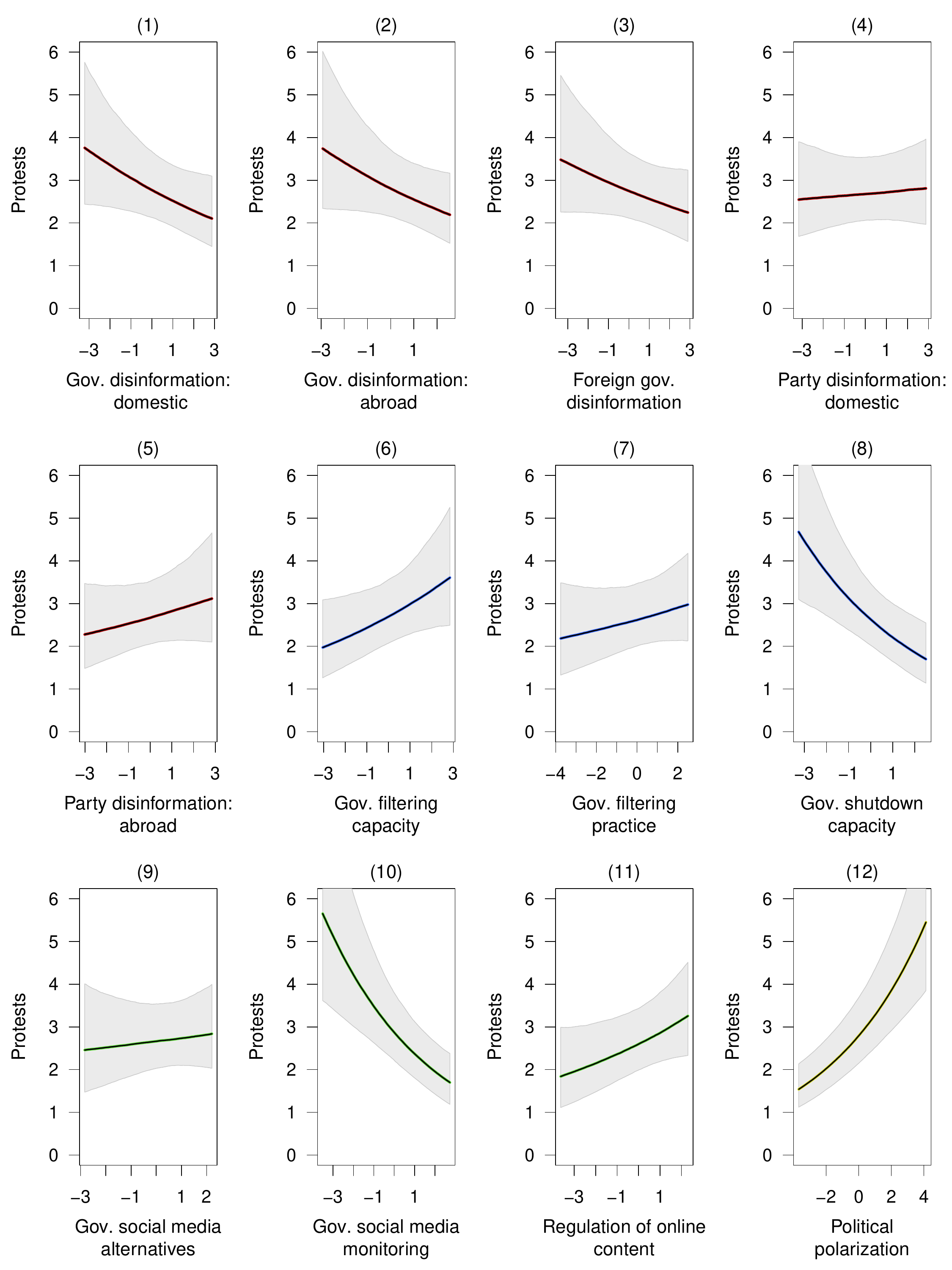}
\caption{Conditional Effects: full sample (Model 3, excluding control variables and random
  effects, 95\% CIs)}
\label{fig: cond}
\end{figure*}

\begin{figure*}[p!]
\centering
\includegraphics[width=\linewidth, height=18cm]{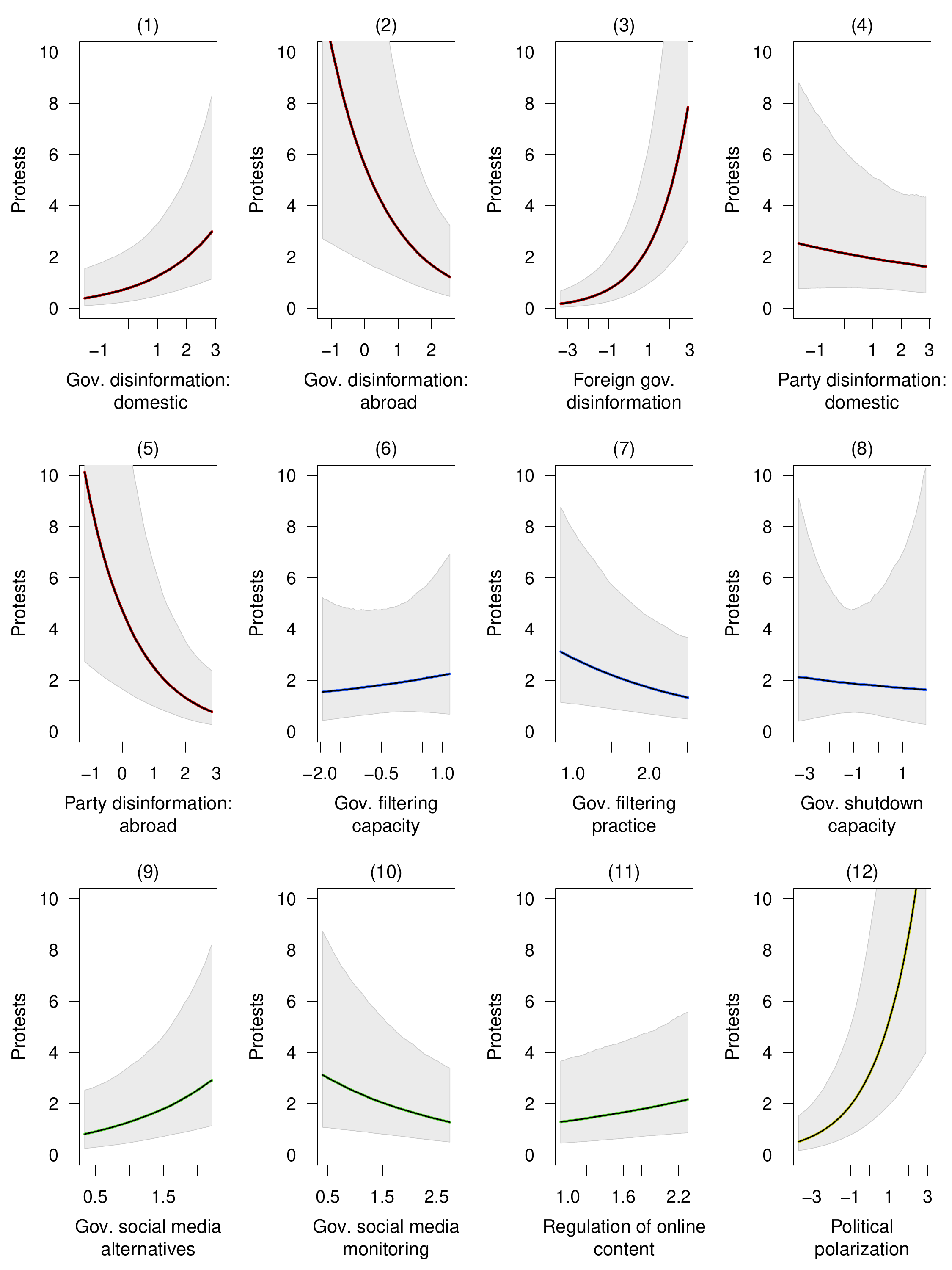}
\caption{Conditional Effects: EEC sample (Model 3, excluding control variables and random
  effects, 95\% CIs)}
\label{fig: cond eec}
\end{figure*}

\begin{table*}[th!b]
\centering
\caption{Regression Results in Two Polarization Regimes (coefficients)}
\label{tab: reg split}
\begin{tabular}{lrrcrr}
\toprule
& \multicolumn{2}{c}{Full Sample}
&& \multicolumn{2}{c}{EEC Sample} \\
\cmidrule{2-6}
%
%& \multicolumn{2}{c}{Polarization} && \multicolumn{2}{c}{Polarization} \\
\cmidrule{2-3} \cmidrule{5-6}
& Low polar. & High polar. && Low polar. & High polar. \\
\hline
Intercept & $-0.98$ & $-4.84$ && $1.79$ & $-18.46$ \\
\cmidrule{1-1}
Gov.~disinformation: domestic & $0.41$ & $-0.35$ && $-0.14$ & $0.24$ \\
Gov.~disinformation: abroad & $-0.47$ & $0.04$ && $0.58$ & $-0.48$ \\
Foreign gov.~disinformation & $0.53$ & $-0.28$ && $< 0.01$ & $0.53$ \\
\cmidrule{1-1}
Party disinformation: domestic & $-0.47$ & $0.16$ && $-1.25$ & $0.48$ \\
Party disinformation: abroad & $0.51$ & $-0.29$ && $-1.06$ & $-0.98$ \\
\cmidrule{1-1}
Gov.~filtering capacity & $-0.06$ & $0.15$ && $0.15$ & $-0.69$ \\
Gov.~filtering practice & $-0.18$ & $0.17$ && $0.68$ & $0.07$ \\
Gov.~shutdown capacity & $-0.95$ & $-0.03$ && $0.32$ & $0.82$ \\
Gov.~social media alternatives & $0.29$ & $-0.04$ && $-0.65$ & $0.55$ \\
Gov.~social media monitoring & $-0.60$ & $-0.07$ && $-0.21$ & $-0.38$ \\
Regulation of online content & $-0.71$ & $0.36$ && $-0.25$ & $-0.45$ \\
\cmidrule{1-1}
National online media & $0.16$ & $-0.03$ && $-0.11$ & $-0.01$ \\
$\ln(\textmd{population})$ & $0.26$ & $0.46$ && $0.57$ & $0.88$ \\
$\ln(\textmd{GDP per capita})$ & $-0.23$ & $-0.20$ && $-0.59$ & $0.67$ \\
\hline
Std.~dev.~years (intercepts) & $0.21$ & $0.26$ && $0.57$ & $0.47$ \\
Std.~dev.~countries (intercepts) & $1.83$ & $0.99$ && $2.40$ & $2.10$ \\
\bottomrule
\end{tabular}
\end{table*}

First, a surprise is immediately present; in all three models, all of the
coefficients for the governmental disinformation variables have negative signs
in the full sample containing all $n = 125$ countries. The same applies for the
two party disinformation variables, which, however, have coefficients with
positive signs in the third model. The magnitudes of the coefficients for the
disinformation variables are also relatively modest. In contrast, in the EEC
sample all of the disinformation variables align with prior expectations; in the
third full model, for instance, positive signs are present for coefficients
measuring governmental propagation of disinformation to domestic audiences,
disinformation spread by foreign governments, and domestic disinformation
disseminated by political parties and major political figures. Negative signs,
in turn, are present for disinformation exported to foreign audiences, whether
done by governmental bodies or political parties. Thus, it seems that if
  a government or a political party in the EEC region is engaged in the
  dissemination of disinformation to foreign audiences, there is less domestic
  political protests in the given country. While it is difficult to interpret
such negative effects upon the protest counts, counter-disinformation by
national security organizations may offer a partial
explanation~\cite[cf.][]{Nakissa20, Ruohonen21GOODIT}. The coefficients are also
relatively large in magnitude for all disinformation variables in the third
model for the EEC sample. The discrepancy between the two samples becomes more
evident when looking at the conditional effects shown in Figs.~\ref{fig: cond}
and \ref{fig: cond eec}. As can be seen, the effects in the EEC sample are
indeed forceful.

Second, similar mixed results are present for the policy variables. In the full
sample the coefficients have negative signs for the variable on the government
capacity for Internet shutdowns. The negative effects are also large when
compared to rest of the policy variables~(see Fig.~\ref{fig: cond}). This
observation can be interpreted to imply that such shutdown measures are
effective in decreasing offline protests. Although such measures are highly
controversial, the causal effect in itself seems logical. In contrast, there is
also a strong negative effect for the variable measuring governmental online
monitoring of political content in social media. By recalling the scale of this
variable, it seems that the absence of such monitoring decreases offline protest
counts, which seems somewhat illogical. Similar observations can be made from
the EEC estimates, although with this sample the conditional effects are small
in magnitude (see Fig.~\ref{fig: cond eec}). Such small effects could be used to
argue that controversial measures such as NetzDG are not effective in reducing
offline protests. Neither does the capacity for Internet shutdowns decrease the
offline protest counts in the EEC sample.

Last but not least, there is a substantial positive effect of political
polarization upon the protest counts. The result is hardly surprising as such;
the more there is polarization, the more there are protests---the logic is
sound. But what remains unclear is whether polarization further mediates the
effects of the disinformation variables particularly in the full sample. This is
the argument recently put forward by \citet{Piazza21} with respect to offline
violence. To examine such meditation, explicit interactions are cumbersome to
implement because there are five disinformation variables. Thus, a simple
computational experiment was carried out instead by estimating the third model
(excluding the regime change dummy variable) in subsamples divided by the median
of the polarization variable.

The results from the sample split estimations are shown in Table~\ref{tab: reg
  split}. When looking at the coefficients and their signs, there indeed exist
differences in terms of the disinformation variables. In the split full sample
relatively large coefficient magnitudes are present in the low polarization
regime within which governmental disinformation to domestic audiences and
foreign government disinformation have coefficients with positive signs. The
opposite holds in the high polarization regime. Thus, the results are still
difficult to interpret for the full sample. In the EEC sample, however, the
previously noted results seem to refer to countries in the high polarization
regime.\footnote{~High polarization countries in the EEC sample include Poland,
  France, Germany, Italy, the Netherlands, Spain, Bulgaria, Croatia, Cyprus,
  Estonia, Finland, Greece, Luxembourg, Romania, Slovenia, and Hungary.} For
these countries, disinformation disseminated to domestic audiences, whether by
governments, political parties, or politicians, tends to increase the protest
counts. All in all, it can be concluded that polarization indeed seems to
meditate the effects of disinformation on protests, although this proxy effect
is neither straightforward to model nor to interpret.

\section{Conclusion}\label{sec: conclusion}

This paper evaluated the effect of online disinformation and other digital
phenomena upon offline political protests in a time-series cross-sectional
sample of $125$ countries between 2000 and 2019. Based Bayesian multi-level
regression analysis, the hypothesis contested holds; there is a statistical
relation between disinformation and protests. However, in the full sample this
relation only becomes visible once political polarization is taken into account;
polarization itself has a strong tendency to increase protests no matter the
sample used. In contrast, the results are clearer, aligning with prior
expectations in the sample of EEC countries; disinformation propagated by both
domestic and foreign governments tends to increase protests. Though, a proxy
effect with political polarization seems to be again present; these effects
apply particularly to countries experiencing high levels of polarization. While
the hypothesis holds, the interpretation is not straightforward.

As for the other independent variables, a government's capacity to shutdown the
Internet has a notable negative effect upon the protest counts in the full
sample of countries. This effect seems logical. If offline political protests
are planned and coordinated online, as is claimed in the literature, shutting
down the Internet obviously decreases protests. Although it could be
  reasoned that governmental social media monitoring would have a deterrence
  effect upon the planning and coordination activities, this line of reasoning
  does not hold. In fact, the absence of governmental monitoring of political
  content in social media seems to paradoxically decrease the offline protest
  counts. Furthermore, in the EEC sample a similar strong effect is absent.
Also previous results have been rather mixed~\citep{Mercea11}. In other words,
online surveillance may either increase or decrease participation particularly
in high-risk offline protests. More generally, according to \citet{Fuchs18},
disinformation, hate speech, privacy, surveillance, and related online phenomena
also constitute one dimension that is shaping the current political activity on
the traditional left-right axis.

A potential factor explaining the diverging results between the two samples is
rooted in the so-called digital divide, which has long been used to describe
societal aspects of information technology between developing and developed
countries~\citep{Rogers01}. There are no particular reasons to suspect that
disinformation and its tenets would be different in this regard. Indeed, the
mixed results can be interpreted also to reflect the existing Western and
English language biases in disinformation research~\citep{Righetti21}. A further
point can be made about construct validity of some of the disinformation
variables supplied in the \text{V-Dem} dataset.

Although information is always both outwardly and inwardly
directed~\citep{Lin20}, it is not entirely clear whether the separation between
domestic and foreign audiences makes sense theoretically. Disinformation, like
propaganda, is often designed for both audiences, and separating the two can be
difficult~\citep[cf.][]{Ruohonen21GOODIT}. Nor is it clear how robust the
protest dataset is for the paper's purposes. Coding protests based on media
articles is subject to known biases, among these the tendency of newspapers to
only report long-lasting or otherwise major
protests~\citep[pp.~37--42]{Almeida19}. In contrast---thanks to social media and
the Internet's overwhelming online advertising business, online disinformation
may be targeted to highly specific groups who are susceptible to provocations
seeking to prompt offline protests. Finally, as contemplated by
\citet{ChangPark20}, a reverse causality may apply; people who participate in
offline protests may turn online in order to continue their protesting. A
similar reasoning may apply to disinformation dissemination.

In addition to patching these limitations, three paths seem important for
further research. First, the meditation effects need further examination, but
complex structural equation models are not necessarily the right tool for the
task because also the longitudinal dimension needs to be accounted for. Second,
further theoretical work is required to distinguish the state-centric viewpoint
to disinformation from other viewpoints. The task is not easy, especially when
considering that theoretical work in disinformation research has been extremely
limited, mainly revolving around the terminological confusion. Finally, third,
more policy-oriented research is required to gain practical relevance. As was
noted in Section~\ref{sec: independent variables}, numerous laws and other
policy responses have already been enacted particularly in Europe. Research,
however, lacks behind; there are not many studies that could reliably inform
policy-making. The present paper is not an exception.

\section*{Funding}

This paper was funded by the Strategic Research Council at the Academy of Finland (grant number~327391). There is a conflict of interest with any other researcher funded by the same grant.

\section*{Ethical Approval}

Not applicable.

\section*{Informed Consent}

Not applicable.

\section*{Author's Contribution}

Not applicable.

\section*{Conflict of Interest}

There is a conflict of interest with other researchers funded by the grant
(no. 327391) from the Strategic Research Council at the Academy of Finland.

\section*{Data Availability}

The datasets used in this paper are publicly available online; please refer to \citet{VDem21a, VDem21b}, \citet{Massmob16a}, and \citet{WorldBank21b,WorldBank21d}.

\balance
\bibliographystyle{apalike}
%\bibliography{media}

\end{document}